\documentclass[a4paper]{IEEEtran}[2015/08/26]

\usepackage{pbalance}

\usepackage{upquote}

\usepackage[ngerman,main=english]{babel}
\addto\extrasenglish{\languageshorthands{ngerman}\useshorthands{"}}

\usepackage[hyphens]{url}

\makeatletter
\g@addto@macro{\UrlBreaks}{\UrlOrds}
\makeatother

\usepackage[zerostyle=b,scaled=.75]{newtxtt}

\usepackage[T1]{fontenc}

\usepackage[
  babel=true, 
  expansion=alltext,
  protrusion=alltext-nott, 
  final 
]{microtype}

\DisableLigatures{encoding = T1, family = tt* }

\usepackage{graphicx}

\usepackage{diagbox}

\usepackage{xcolor}

\usepackage[autostyle=true]{csquotes}

\defineshorthand{"`}{\openautoquote}
\defineshorthand{"'}{\closeautoquote}

\usepackage{booktabs}

\usepackage{paralist}

\usepackage[%
  square,        
  comma,         
  numbers,       
  sort&compress 
]{natbib}

\usepackage{stfloats}
\fnbelowfloat

\usepackage[group-minimum-digits=4,per-mode=fraction]{siunitx}
\addto\extrasgerman{\sisetup{locale = DE}}

\usepackage{amsmath}
\usepackage{amssymb}
\usepackage{rotating}

\usepackage{hyperref}

\hypersetup{
  hidelinks,
  colorlinks=true,
  allcolors=black,
  pdfstartview=Fit,
  breaklinks=true
}

\usepackage[all]{hypcap}

\usepackage[caption=false,font=footnotesize]{subfig}

\usepackage[capitalise,nameinlink,noabbrev]{cleveref}

\crefname{listing}{Listing}{Listings}
\Crefname{listing}{Listing}{Listings}
\crefname{lstlisting}{Listing}{Listings}
\Crefname{lstlisting}{Listing}{Listings}

\DeclareFontFamily{U}{MnSymbolC}{}
\DeclareSymbolFont{MnSyC}{U}{MnSymbolC}{m}{n}
\DeclareFontShape{U}{MnSymbolC}{m}{n}{
  <-6>    MnSymbolC5
  <6-7>   MnSymbolC6
  <7-8>   MnSymbolC7
  <8-9>   MnSymbolC8
  <9-10>  MnSymbolC9
  <10-12> MnSymbolC10
  <12->   MnSymbolC12%
}{}
\DeclareMathSymbol{\powerset}{\mathord}{MnSyC}{180}

\usepackage{xspace}

\makeatletter
\newcommand{\hydash}{\penalty\@M-\hskip\z@skip}
\makeatother

\input glyphtounicode
\makeatletter
\newcommand*\titleheader[1]{\gdef\@titleheader{#1}}
\AtBeginDocument{%
  \let\st@red@title\@title
  \def\@title{%
    \bgroup\normalfont\large\centering\@titleheader\par\egroup
    \vskip.5em\st@red@title}
}
\makeatother

\titleheader{\emph{Draft. The final version will appear in the proceedings of the\\ International Symposium on Leveraging Applications of Formal Methods (ISoLA) 2022.}}

\title{Capturing Dependencies within Machine Learning via a Formal Process Model}

\author{%
\vspace*{2mm}
\IEEEauthorblockN{Fabian Ritz\IEEEauthorrefmark{1}, Thomy Phan\IEEEauthorrefmark{1}, Andreas Sedlmeier\IEEEauthorrefmark{1}, Philipp Altmann\IEEEauthorrefmark{1}, Jan Wieghardt\IEEEauthorrefmark{2},\\Reiner Schmid\IEEEauthorrefmark{2}, Horst Sauer\IEEEauthorrefmark{2}, Cornel Klein\IEEEauthorrefmark{2}, Claudia Linnhoff-Popien\IEEEauthorrefmark{1} and Thomas Gabor\IEEEauthorrefmark{1}}\\
\vspace*{2mm}
\IEEEauthorblockA{\IEEEauthorrefmark{1}Mobile and Distributed Systems Group, LMU Munich, Germany}\\
\IEEEauthorblockA{\IEEEauthorrefmark{2}Technology, Siemens AG, Germany}\\
\vspace*{2mm}
\IEEEauthorblockA{Corresponding authors: \{fabian.ritz,thomas.gabor\}@ifi.lmu.de}
\vspace*{-4mm}
}

\begin{document}


\maketitle
\thispagestyle{empty}

\begin{abstract}
The development of Machine Learning (ML) models is more than just a special case of software development (SD):
ML models acquire properties and fulfill requirements even without direct human interaction in a seemingly uncontrollable manner.
Nonetheless, the underlying processes can be described in a formal way.
We define a comprehensive SD process model for ML that encompasses most tasks and artifacts described in the literature in a consistent way.
In addition to the production of the necessary artifacts, we also focus on generating and validating fitting descriptions in the form of specifications.
We stress the importance of further evolving the ML model throughout its life-cycle even after initial training and testing.
Thus, we provide various interaction points with standard SD processes in which ML often is an encapsulated task.
Further, our SD process model allows to formulate ML as a (meta-) optimization problem.
If automated rigorously, it can be used to realize self-adaptive autonomous systems.
Finally, our SD process model features a description of time that allows to reason about the progress within ML development processes.
This might lead to further applications of formal methods within the field of ML.
\end{abstract}


%
\IEEEpeerreviewmaketitle

\section{Introduction}\label{sec:introduction}
\begin{figure}
    \centering
	\includegraphics[width=0.48\textwidth]{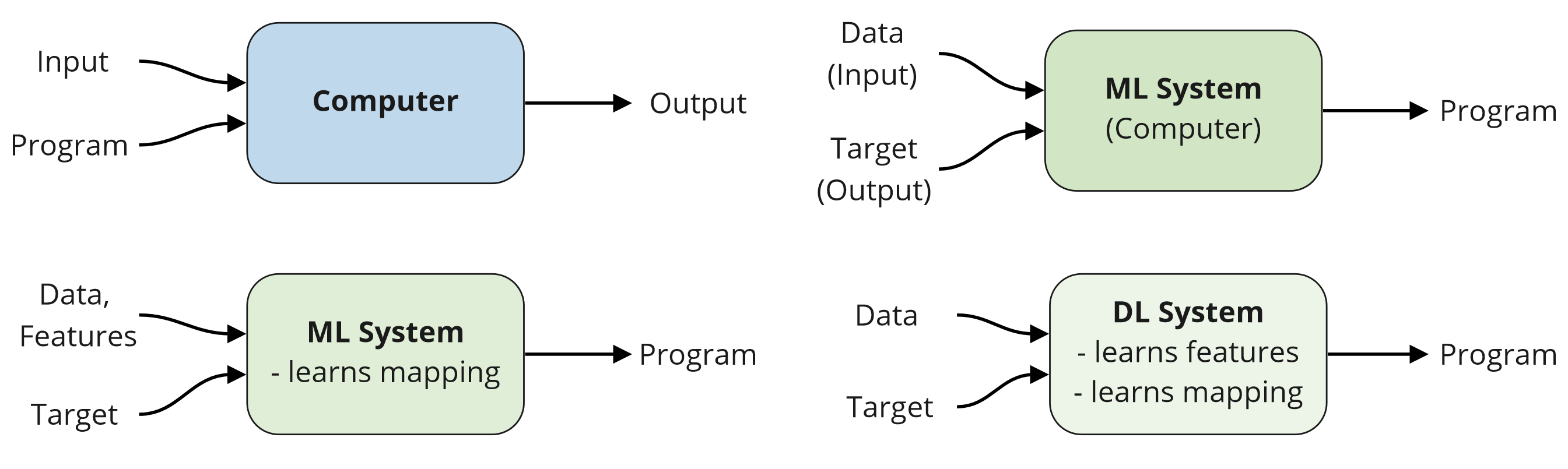}
	\caption{Comparison of traditional software development (upper left) and ML software development (upper right). More specifically, DL systems (lower right) are a special type of ML systems (lower left): DL systems automatically learn the relevant input features and how to map these features to outputs. While this reduces the effort previously required to define features and mapping logic, it makes it more difficult to understand the rules by which DL systems make decisions. Illustration based on \cite{gir21}.}
	\label{fig:cs_ml_dl_comparison}
\end{figure}

In recent software systems, functionality is often provided by \textit{Machine Learning} (ML) components, e.g. for pattern recognition, video game play, robotics, protein folding, or weather forecasting. 
ML infers a statistical model from data, instead of being programmed explicitly.
In this work, we focus on the usage of \textit{Deep Learning} (DL) techniques, which presently are the most commonly used approaches.
Figure~\ref{fig:cs_ml_dl_comparison} (based on \cite{gir21}) gives a high-level overview how traditional software systems and ML systems are typically developed. 
\textit{Software Engineering} (SE) for ML systems is an emerging research area with an increasing number of published studies since 2018 \cite{mbf+21,gir21}.
In practice, software companies face a plethora of challenges related to data quality, design methods and processes, performance of ML models as well as deployment and compliance \cite{bch20,mbf+21,gir21}.
Thus, there is a need for tools, techniques, and structured engineering approaches to construct and evolve these systems.

\begin{figure*}
    \centering
	\includegraphics[width=\textwidth]{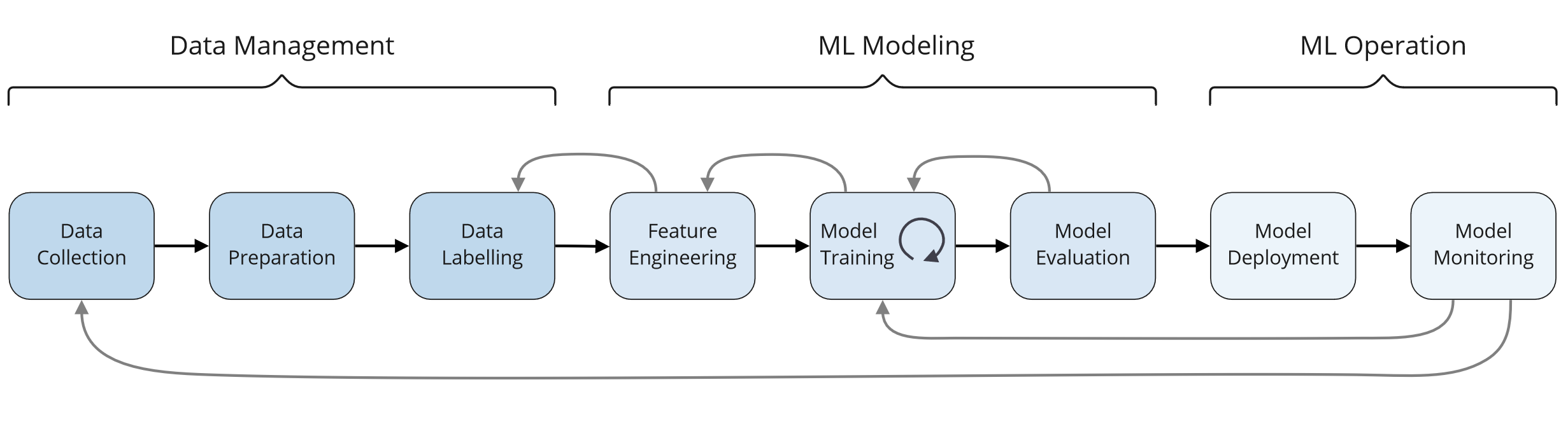}
	\caption{Typical ML development workflow stages with activites and feedback flow. Illustration based on \cite{lcb20}.}
	\label{fig:ml_workflow}
\end{figure*}

\textit{Software Development} (SD) for ML systems typically starts with the management of data, e.g. collection and labeling.
Then, repeated train-test cycles of the ML model are performed, e.g. for hyper-parameter tuning, until the expected results are obtained.
After that, the ML model is deployed and monitored during operation.
Feedback from operation is typically used to re-train (patch) an existing ML model or to extend the data-sets.
This process is called \textit{ML workflow} and is visualized in Figure~\ref{fig:ml_workflow} (based on \cite{lcb20}).
So far, a number of ML workflows and life-cycle models have been constructed in an ad-hoc manner based on early experiences of large software companies, e.g. reported by IBM \cite{asx+20}, Microsoft \cite{abb+19}, or SAP \cite{rrk+19}.
The respective case studies focus strongly on the ML workflow but little on the integration with existing SE processes and tools, thus not covering the entire SD process.

Then, \textit{MLOps}~\cite{lcb20,kkh22} emerged as an end-to-end ML development paradigm (see Figure~\ref{fig:mldevops}).
MLOps combines the DevOps process, i.e. fast development iterations and continuous delivery of software changes, with the ML workflow.
Further collaboration between industry and academia resulted in the development of the \textit{CRISP-ML(Q)} life-cycle model~\cite{sbd+21}, which additionally contains technical tasks for quality assurance.
Yet, we found existing MLOps process models and CRISP-ML(Q) lacking a clear view on the dependencies between activities and the involved artifacts.

Another key aspect is automation.
Presently, software updates (or patches) are still often performed manually by human developers or engineers.
To minimize the impact of dynamic changes, agile SD paradigms such as MLOps perform smaller, faster updates through cross-functional collaboration and high automation.
Ultimately, however, we might want to minimize the amount of human effort spent during SD.
ML systems are able to adapt to changes via generalization and can be re-trained with little human effort.
One step further, \textit{Auto-ML} \cite{hzc21} is a popular tool to automate single steps of the ML workflow: the search for a suitable ML model architecture and the adjustment of the according hyper-parameters.
The next advancement would be to enable automated optimization spanning over multiple steps of SD processes.
One example would be to autonomously decide when to re-train an ML model that is already in production using newly collected data.
Conceptually similar approaches already exist in the field of (collective) autonomic systems~\cite{whk+15}.
In practice, this would require a tight integration of quality assurance activities.
Engineering trustworthy ML systems is an ongoing challenge since it is notoriously difficult to understand the rules by which these systems make decisions~\cite{lpk21}.
Adding self-optimization on SD process level will most likely increase the complexity.

To tackle these challenges, we propose a formal process model for ML that can be mapped to existing SD processes.
Our process model is based on practical findings from various cooperations with industry partners on Multi-Agent Reinforcement Learning~\cite{pga+20,rpm+22}\footnote[1]{\url{https://www.siemens.com}} and Anomaly Detection problems~\cite{mir+21}\footnote[2]{\url{https://www.swm.de}}. 
It encompasses the majority of ML-specific tasks and artifacts described in the literature in a consistent way.
Further, it allows for automation spanning over multiple steps of SD processes.
It is not restricted to certain feedback loops and supports self-optimization.
If automated rigorously, it can be used realize self-adaptive, autonomous systems.

The remainder of the paper is structured as follows:
In Section~\ref{sec:related_work}, we provide an overview of related work regarding SE for (self-) adaptive systems (Section~\ref{sec:se4as}), SE for ML systems (Section~\ref{sec:se4ml}) and fundamental challenges of ML that could be alleviated through SE (Section~\ref{sec:ml_challenges}).
In the following Section~\ref{sec:formal_process_model}, we visualize (Section~\ref{sec:visualization}) and describe (Section~\ref{sec:description_of_elements}) our process model.
We then present a proof of concept (Section~\ref{sec:poc}) and provide a brief formalization of our process model (Section~\ref{sec:formalization}).
Finally, in Section~\ref{sec:discussion}, we conclude with a summary of strengths and limitations and provide an outlook for future work.

\section{Background and Related Work}\label{sec:related_work}
Designing systems that autonomously adapt to dynamic problems is no new trend.
What has changed with the emergence of modern DL techniques is the ability to handle greater state and action spaces.
Still, it is often impossible to design explicit control systems that directly adapt the exactly right parameters to changed conditions (\textit{parameter adaptation}) because there are too many and potentially unknown relevant parameters.
However, it is possible to design systems which cause change on a higher level to meet the new conditions.
In the literature, this concept is referred to as \textit{compositional}~\cite{msk+04} or \textit{architecture-based} adaptation~\cite{gsc09}.
Following this concept, we classify a system as \textit{self-adaptative} if it autonomously changes to meet (at least) one of the following three aspects~\cite{gei08}:
\begin{enumerate}
\item The implementation of one component is replaced by another.
\item The structure of the system changes, i.e. the relations between components change, components are added or removed.
\item The distribution of the system components changes without modification of the logical system structure, e.g. components can migrate.
\end{enumerate}
Through generalization, ML systems are capable of parameter adaptation out-of-the-box and they can further be used to realize self-adaptive systems.
It is already common to re-train ML models once new data is gathered during operation and then replace the deployed model with the re-trained one.
Once such work-flows are fully automated, such systems are self-adaptive as per the above definition.

One key to engineer such systems will be \textit{Verification} \& \textit{Validation} (V\&V) activities, which shall build quality into a product during the life-cycle~\cite{ieee12}.
Verification shall ensure that the system is built correctly in the sense that the results of an activity meet the specifications imposed on them in previous activities~\cite{bf14}.
Validation shall ensure that the right product is built.
That is, the product fulfills its specific intended purpose~\cite{bf14}.
Related to this, Gabor et al. analyzed the impact of self-adapting ML systems on SE, focusing on quality assurance \cite{gsp+20}.
They provide a general SD process description related to ML and embed it into a formal framework for the analysis of adaptivity.
The central insight is that testing must also adapt to keep up with the capabilities of the ML system under test.
In this paper, we build a process model around the ML life-cycle and provide insights about the interplay of SD activities and artifacts.

\subsection{SE for (Self-)Adaptive Systems}\label{sec:se4as}
Researchers and practitioners have begun tackling the SE challenges of self-adaptation prior to the latest advances in DL.
Influencing many later following approaches, Sinreich \cite{s06} proposed a high-level architectural blueprint to assist in delivering autonomous computing in phases.
Autonomous computing aims to de-couple (industrial) IT system management activities and software products through the SD cycle.
The proposed architecture reinforces the usage of intelligent control loop implementations to \textit{Monitor, Analyze, Plan and Execute, leveraging Knowledge} (MAPE-K) of the environment.
The blueprint provides instructions on how to architect and implement the knowledge bases, the sensors, the actuators and the phases.
It also outlines how to compose autonomous elements to orchestrate self-management of the involved systems.
As of today, the paradigms of autonomous computing have spread from IT system management towards a broad field of problems, the most prominent one being autonomous driving.
Our process model is built to specifically assist the development of systems that are based on DL techniques.

To augment development processes to account for even more complex systems, Bernon et al. later proposed ADELFE~\cite{bcg+04,bcg+05}.
Built upon RUP \cite{k00}, ADELFE provides tools for various tasks of software design.
From a scientific view, ADELFE is based on the theory of adaptive Multi-Agent systems: they are used to derive a set of stereotypes for components to ease modeling.
Our process model also supports this architectural basis (among others), but does not require it.

In the subsequent ASCENS project \cite{whk+15}, a life-cycle model formalizes the interplay between human developers and autonomous adaptation.
It features separate states for the development progress of each respective feedback cycle.
Traditional SD tasks, self-adaptation, self-monitoring and self-awareness are regarded as equally powerful contributing mechanisms.
This yields a flexible and general model of an engineering process but does not define a clear order of tasks.
The underlying \textit{Ensemble Development Life Cyle} (EDLC) \cite{hkp+15} covers the complete software life cycle and provides mechanisms for enabling system changes at runtime.
Yet, interaction points between traditional SD and self-adaptation are only loosely defined.
Recently, AIDL \cite{wb22} specialized the EDLC to the construction of autonomous policies using Planning and Reinforcement Learning techniques.
Overall, the ASCENS approach emphasizes formal verification.
Correct behavior shall be proven even for adapted parts of the system.
Analogous to SCoE \cite{gsp+20}, we agree a strong effort for testing is necessary when adaptation comes into play.
Our process model was built with self-adaptation in mind and allows for a seamless integration of V\&V activities and respective feedback loops at any point in time.

\subsection{SE for ML Systems}\label{sec:se4ml}
\begin{figure*}
    \centering
	\includegraphics[width=\textwidth]{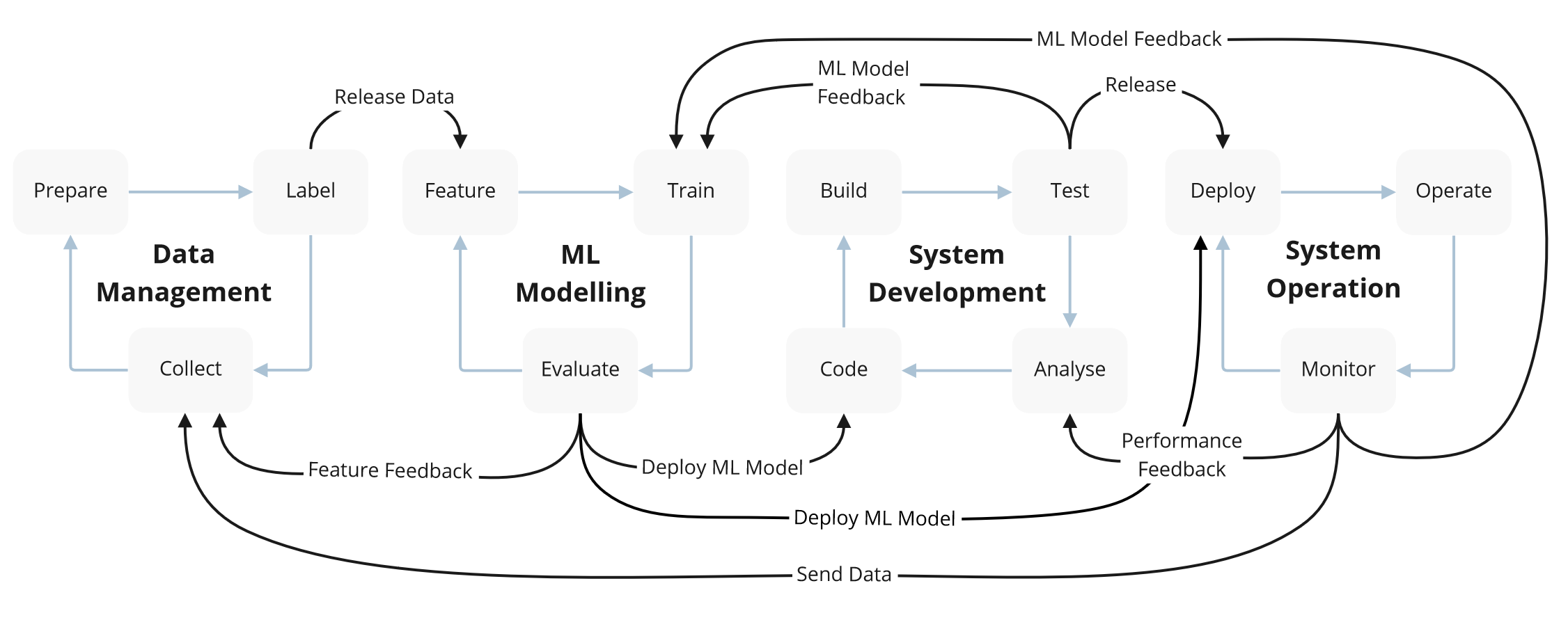}
	\caption{Illustration of an MLOPs process model based on \cite{lcb20}. The ML workflow is integrated into DevOps by adding \textit{Data Management} and \textit{ML Modeling} to the existing \textit{Development} and \textit{Operations} processes and (feedback) transitions.}
	\label{fig:mldevops}
\end{figure*}
Regarding SE for ML systems during the latest advances in DL, literature has been systematically reviewed several times~\cite{mbf+21,gir21}.
A common conclusion is that all SE aspects of engineering ML systems are affected the non-deterministic nature of ML (even slight changes in the setup can have a drastic impact) but none of the SE aspects have a mature set of tools and techniques yet to tackle this.
According to the authors, the reported challenges are difficult to classify using the established SWEBOK Knowledge Areas~\cite{bf14} since they are strongly tied to the problem domain.
Our process model explicitly considers V\&V to ensure quality of ML software independent of the problem domain.

To identify the main SE challenges in developing ML components, Lwakatare et al.~\cite{lrb+19} evaluated a number of empirical case studies.
They report a high-level taxonomy of the usage of ML components in industrial settings and map identified challenges to one of four ML life-cycle phases (\textit{assemble data set}, \textit{create model}, \textit{train and evaluate model}, \textit{deploy model}).
These life-cycle phases can be mapped to our process model.
Building upon the taxonomy of Lwakatare et al., Bosch et al.~\cite{bch20} propose a research agenda including autonomously improving systems.
Besides data management related topics, the elements of the research agenda can be mapped to our process model.

During the recent emergence of \textit{MLOps} (an overview is given in~\cite{kkh22}), Lwaka-tare et al.~\cite{lcb20} proposed a precise and clear variant that integrates the ML workflow into DevOps by adding the \textit{Data Management} and \textit{ML Modeling} to \textit{Development} and \textit{Operations}, effectively expanding the ML workflow to an end-to-end SD process (see Figure~\ref{fig:mldevops}).
The resulting process model aims for automation at all stages and enables iterative development through fast feedback flow.
Building on MLOps, \textit{CRISP-ML(Q)} \cite{sbd+21} describes six phases ranging from defining the scope to maintaining the deployed ML application. 
Challenges in the ML development are identified in the form of risks.
Special attention is drawn to the last phase, as they state that a model running in changing real-time environments would require close monitoring and maintenance to reduce performance degradation over time.
Compared to MLOps, CRISP-ML(Q) additionally considers business understanding and ties it closely to data management.
However, we found MLOps and CRISP-ML(Q) lacking a precise view on the dependencies between activities and artifacts, which our process model tries to accomplish.

Finally, Watanabe et al.~\cite{wws+19} provide a preliminary systematic literature review of general ML system development processes.
It summarizes typical phases in the ML system development and provides a list of frequently described practices and a list of adopted traditional SD practices. 
The phases in our process model share common ground with the granular phases described in this literature review.
Generally, we provide various interaction points with standard SD processes (in which ML often is an encapsulated task) by evolving the ML model throughout its life-cycle after initial training and testing.

\subsection{Fundamental ML Challenges}\label{sec:ml_challenges}
So far, we considered emerging issues of SE when incorporating ML components.
Vice versa, ML faces some fundamental challenges (besides the current technical difficulties already mentioned) that might not be solved through technical improvements alone but might require to be addressed through SE.
One fundamental challenge of ML is \textit{explainability}, i.e. methods and techniques in the application of AI such that the decisions leading to the solutions can be understood (at least) by human experts \cite{lpk21}.
Although some approaches address explainability, e.g. by gaining post-hoc insights about the predictions \cite{rsg16}, and proposals were made to use techniques that are explainable per default \cite{r19}, most current ML models remain black boxes in the sense that even ML engineers cannot always explain why the ML models produce a certain output.
Though each individual parameter within an ML model can be analyzed technically, this does not answer the relevant questions about causal relationships.
This can cause legal problems in areas such as healthcare or criminal justice.
More importantly, it is notoriously difficult to validate and verify these ML models, both from an engineering and an SD process point of view.
Systematic guidance through the ML life-cycle to enable trustworthiness of the ML models would help.

Another fundamental challenge of ML is \textit{efficiency}.
In general, modern ML relies on DL techniques.
The predictive performance of these models scales with their size, which requires the available of more training data and more computational power to optimize the risen amount of parameters.
ML model complexity and the computational effort required to train these models grew exponentially during the last years \cite{sv21}.
Developing such ML models was only possible due to advances in hardware and algorithmic design \cite{hb20}.
Whether further improvements with ML can be achieved this way is uncertain.
Nonetheless, re-training of state-of-the-art ML models requires significant amounts of computational resources and time.
Because ML suffers from the ``changing one thing changes everything'' principle, it may be costly to fix issues in ML models afterwards regardless of what caused the issue in the first place.
Consequently, support from SE to ensure high quality ML models upfront, e.g. through precise requirements and feedback loops, is crucial to effectively use ML in production.

\section{A Process Model to Capture Dependencies within ML}\label{sec:formal_process_model}
SE has given rise to a series of notions to capture the dynamic process of developing software.
Opinions on which notion is the best differ between different groups of people and different periods of time.
As a common ground, we assume that SD can be split into various \textit{phases} with each phase focusing on different sets of \textit{activities}.
We acknowledge that in agile approaches to SE these phases are no longer to be executed in strict sequential order, so that phases may overlap during development.
But we still regard \textit{planning}, \textit{development}, \textit{deployment}, and \textit{operations} to be useful categories to group activities in SD phases.

First of all, we model \textit{human tasks} and \textit{automated procedures} through activities.
They may require different skill sets, rely on different (computational or human) resources, and are driven by different \textit{key factors}.
E.g., the use case analysis should be performed by business analysts and result in some form of (requirements) specification.
Based on that, data scientists can select or gather suitable data for ML training.
Naturally, business analysts and data scientists should collaborate closely.

Second, we model activities to result in \textit{artifacts}, which may then be required for other activities.
In the above example, the choice of data naturally depends on the problem domain, thus relies on a requirements specification.
Consequently, the use case analysis resulting in the requirements specification should be finished first.
Artifacts can have different types such as \textit{data}, \textit{functional descriptions} (e.g. the ML-Model), or \textit{logical statements} (e.g. a specification, a set of hyper-parameters, or a test result indicating that a specific requirement is fulfilled by an ML model with a specific set of hyper-parameters on specific data).

Third, we capture feedback and (self-) optimization through loops that go back to earlier stages of the ML development process.
This way, we take the iterative nature of ML SD into account, which consists of repeated stochastic optimization and empirical validation of the ML model at different development stages.
Thus, the process model is flexible regarding the employed ML development method and considers V\&V aspects.
It supports manually performed hyper-parameter optimization as well as fully automated patch deployment of a model that was re-trained with data gathered during earlier live operation.
Due to the vast number of possibilities, we leave the actual definition of feedback loops open for the respective use case.

All in all, the purpose of our process model is to capture dependencies between activities and artifacts during the ML life-cycle and to close the gaps between existing SD process models and specialized ML training procedures.
\begin{sidewaysfigure*}
    \centering
	\includegraphics[width=\textwidth]{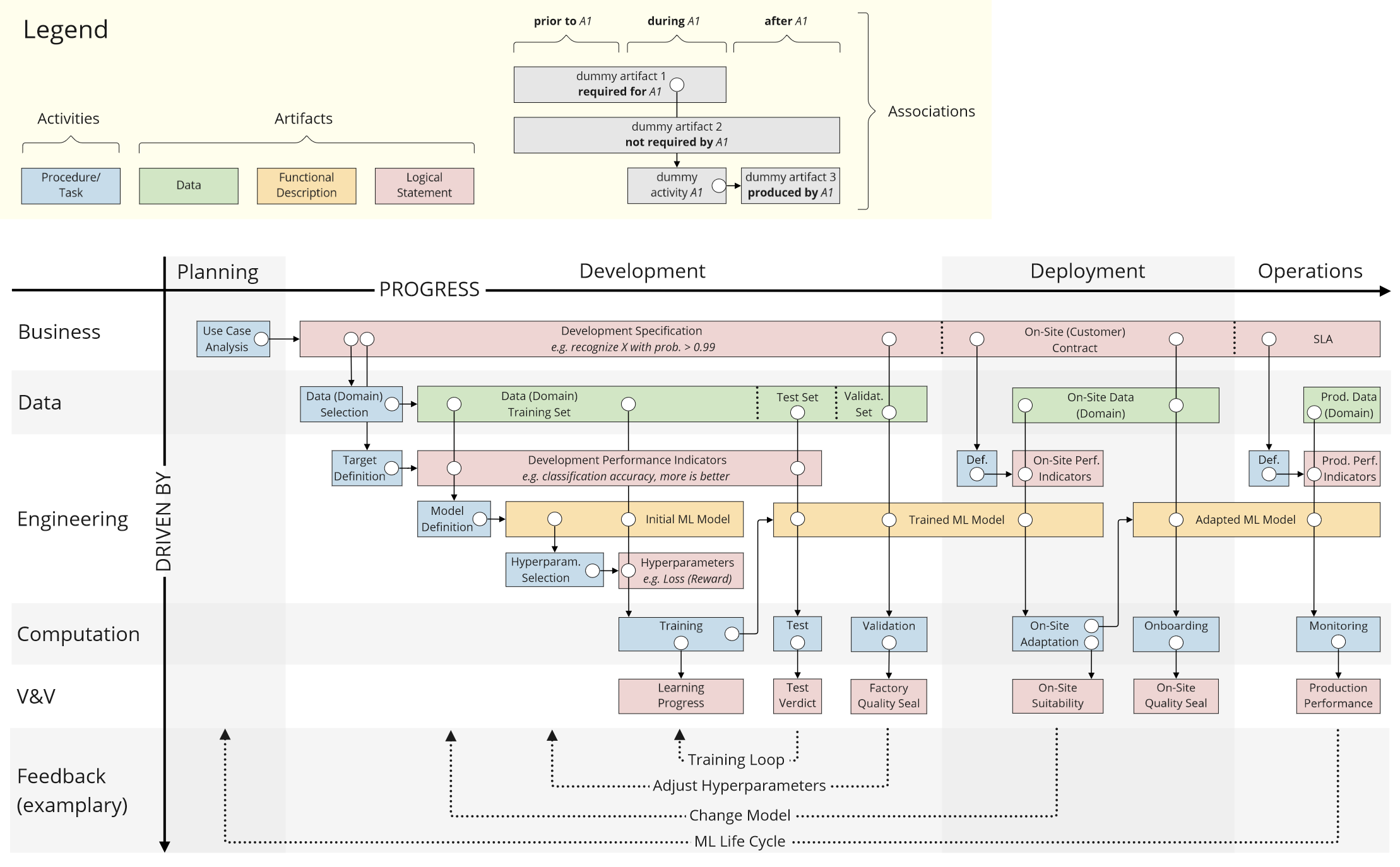}
	\caption{Visualization of the proposed process model for the development of ML software.}
	\label{fig:ml_pipeline}
\end{sidewaysfigure*}

\subsection{Visualization}\label{sec:visualization}
This section provides a visual representation of the proposed process model and explains concept and semantics.
As to be seen in Figure~\ref{fig:ml_pipeline}, the elements are arranged within a temporal dimension (x-axis) and an organizational dimension (y-axis).
Within the temporal dimension, the elements are grouped by four SD phases, which allows a straightforward mapping to existing SD process models.
Further right on the x-axis means later within the development process, i.e. its progress towards completion is higher.
The categories within the organizational dimension are a mixture of business and technical areas.
We refrain from exhaustively defining role names or responsibilities.
Instead, we want to sketch the different key factors that drive the process forward (or limit it).
As mentioned before, the overall categorization of elements within these dimensions is by no means strict and should be adapted to the respective circumstances.

We distinguish between activities, artifacts and associations.
Activities include human tasks and automated procedures.
Artifacts can be data, logical statements or functional descriptions.
Activities always \textit{produce} at least one artifact and can \textit{require} other artifacts.
Associations between artifacts and activities are represented by arrows.
If an activity requires an artifact, a white circle connects that artifact to an arrow whose head ends on the activity.
If an activity produces an artifact, a white circle connects the activity to an arrow whose head ends on that artifact.

For any progress of the development process, there exists (at least) one activity.
For any activity, all activities left to it are considered completed.
Vice versa, all activities right to it are not.
An activity starts if all required artifacts exist.
If this is the case, the respective activities are considered active by default.
Activities are neither connected to other activities, nor require a dedicated trigger.

Please note that we made some trade-offs to ease readability of the visualization.
First, we kept some spacing between elements on the x-axis.
Thus, there are spots without activities (however, there still are associations from or towards activities).
Next, there are no multiplicities on the associations.
Syntactically, it does not matter if multiple artifacts are connected with an activity through one shared or multiple individual arrows. 
Most importantly, the size of the elements is determined by the length of their descriptions and the layout, but does not state anything about their duration in real time.

Finally, there is a number of dashed, annotated arrows in the feedback category of the organizational dimension.
They start once some V\&V artifacts exist and lead back to earlier points in time.
These are examples for different feedback or (self-) optimization loops that define whether an iterative, a monolithic, or a mixture approach is used.
In practice, we expect quality gates to decide, based on the V\&V artifacts, how the SD process continues at certain points.
However, this depends strongly on organizational and legal requirements as well as technical conditions.

\subsection{Description of Activities}\label{sec:description_of_elements}
Our SD process model is built around activities, which we briefly describe in the following (grouped by their respective phase).

\textbf{Planing phase}: We begin with a \textit{use case analysis} activity, which we expect to be driven mainly by business.
The resulting development (requirements) specification defines the goals in natural language.
We acknowledge that this specification may change to some extent during the ML life-cycle, e.g. when the ML model is deployed at the customer's site or system later in the process.
In any case, it is crucial to consider the probabilistic nature of ML when formulating the specification~\cite{mbf+21,gir21}.

\textbf{Development phase}: The first activity here is the \textit{selection (or assembly) of the data set} based on the development specification.
We assume that the resulting data set will usually be split into a training, a testing and validation part, which is indicated by dashed lines.
The bottleneck and driving force here is data management.
We omit activities related to data management and instead assume that suitable data is accessible in some form.
How to construct data sets such that they correspond to the requirements is a challenging problem on its own~\cite{pie+21}. 
In parallel, the \textit{definition of the training target} for the ML model takes place, resulting in development performance indicators that reflect the development specification in a more technical form.
E.g., a requirement may be to correctly recognize specific entities on images with a probability of more than 0.99.
A suitable performance indicator could be prediction recall, where higher values are better.
We consider it important to clearly distinguish between the development specification and the training target (with the respective performance indicators) for two reasons.
The first one is to avoid misconceptions between business analysts and ML experts.
The second one is to enable an SD process controlled curriculum, e.g. to begin training with a subset of the overall specification and once this training target is reached, it is expanded gradually until the overall requirements are met.

The following activity is the \textit{ML model definition}.
Here, the architecture of the ML model is chosen based on the data and the performance indicators.
Then, the \textit{hyper-parameter selection} activity follows, based on the initial ML model.
The hyper-parameters can be algorithm-specific and also include the mathematical optimization target such as the loss in case of \textit{Supervised Learning} (SL) and the reward in case of \textit{Reinforcement Learning} (RL).
Again, decoupling the low-level optimization target from the higher-level training target and the top-level development specification is important:
deriving a suitable loss- or reward-function for a specific training target is a challenging problem on its own \cite{rpm+22}.
Also, separating reward and training target measurements allows to detect reward hacking, a known challenge of RL.
Through decoupling, SD process controlled (self-) optimization can be realized here.

Then, the \textit{training} of the ML model takes place.
The current learning progress can be assessed through the history of the mathematical optimization target (loss or reward) of the ML model on the training data (or training domain in case of RL).
Optionally, the training target can additionally be used to determine the learning progress during training.
However, from a V\&V point of view, the training target should be optimized implicitly through the mathematical optimization target.
In any way, the training target will be assessed during the next, usually periodical activity: \textit{testing} the trained ML model with the development performance indicators on the test data set (or test domain in case of RL).
In practice, we expect different feedback loops from the respective test verdict to different prior activities, especially model definition and hyper-parameter selection, due to the iterative nature of ML training.
This is where Auto-ML~\cite{hzc21} comes into action.
The final activity of the development phase is the \textit{validation} (or evaluation) of the trained ML model against the development specification on the validation data set (or domain in case of RL) resulting in a "factory quality seal".

\textbf{Deployment Phase}: Once the validation activity is passed, we move on by leaving the controlled (maybe simulated) development environment.
The trained ML model faces less controlled conditions, i.e. the individual customer's system or a physical environment.
Thus, the top-level specification may now differ from the one used during development and is referred to as on-site contract.
Most likely, this on-site contract is a subset of the development specification, but it may also contain some additional requirements.
As the requirements and the data (or the domain) are provided by the customer, the first activity here is to \textit{define on-site targets} and performance indicators to take the specialized on-site contract into account.
Then, we expect some \textit{on-site adaptation} of the trained ML model to take place, most likely in the form of additional re-training, followed by a on-boarding (validation) of the adapted ML model which should now fulfill the on-site contract.
If significant specialization is required, on-site adaptation could be extended to a cycle similar to the train-test-cycle during the development phase.
However, we consider it more likely that this is a fast, slim process step (given a thorough initial requirements engineering and training).
The \textit{onboarding} (test) of the adapted ML is the final activity during the deployment phase.
Its result, the "on-site quality seal", states whether the ML model fulfills the provided on-site contract based on the provided data (or domain in case of RL).

\textbf{Operations Phase}: After a successful onboarding, the adapted ML model can be used in production.
We expect the top-level specification to differ from the preceding on-site contract (e.g., to reflect use-case specifics recognized during onboarding), thus referring to it as SLA.
We address this through a \textit{definition of production targets} and the respective performance indicators.
These are the key to meaningful \textit{monitoring} of the ML model on production data, e.g. to detect distributional drift, which can lead to a slow degradation of performance over time.
If no appropriate monitoring is present, such changes may remain undetected and can cause silent failure~\cite{sbd+21}.
Also, identifying situations in which the ML model underperforms is the key for precise feedback used to train future ML models, e.g. through updated data or domain simulations.

\subsection{Proof of Concept}\label{sec:poc}
In this section, we briefly present a practical application of the proposed methodology that was published with a different focus in \cite{rpm+22} as a proof of concept.
Since the process model presented in this paper was created based on experience from this (and similar) cooperations with industry, it naturally fits very well.
Nevertheless, we hope to contribute further practical examples in engineering ML modules with our process model in future work, hopefully not only conducted by us.

\begin{figure*}
    \centering
	\includegraphics[width=\textwidth]{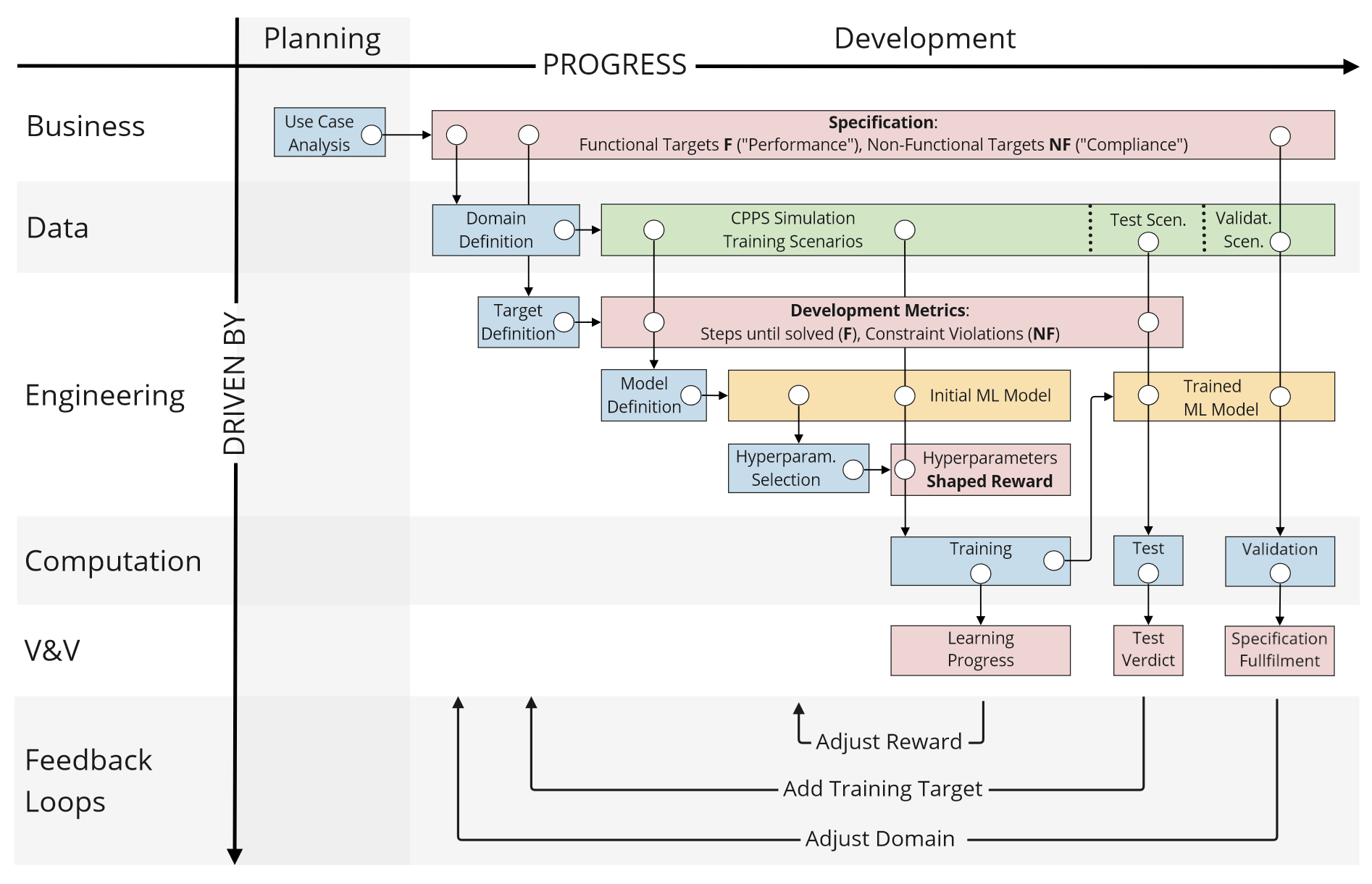}
	\caption{Visualization of a practical application of the proposed proposed process model that was published with a different focus in \cite{rpm+22}. Functional and non-functional requirements were embedded step-wise into the reward function of a cooperative Multi-Agent system. This enabled the RL agents to align with the provided specification.}
	\label{fig:pm_sa-marl}
\end{figure*}

In \cite{rpm+22}, we presented a specialized RL training method that embeds functional and non-functional requirements in step-wise fashion into the reward function of a cooperative Multi-Agent system.
Thus, the RL agents were able to learn to fulfill all aspects of a given specification.
This approach was then compared to naive approaches, which by contrast were only able to fulfill either the functional or the non-functional requirements.
Figure~\ref{fig:pm_sa-marl} visualizes the core of the approach with two feedback loops adjusting the training target and the reward.
The third feedback loop takes into account that the environment simulation could also be adjusted.
Consider that, for example, the desired collaborative task might initially be too difficult and prohibits any learning progress.
In this case, training could start with a small number of simultaneously trained agents and the third feedback loop could gradually increase it, thus creating a curriculum of increasing difficulty.
Although hand-crafted adaptation schedules were used in the approach, this could be realized autonomously in future applications.
The physical counterpart of the environment simulation has not yet been realized, thus there is no deployment or production phase (yet).

\subsection{Formalization}\label{sec:formalization}
In the following, we briefly formalize our process model and sketch the potential we see in the application of formal methods like \textit{Linear Time Logic} (LTL)~\cite{km08}.

The development process $\mathfrak{D}$ is given by a set of elements $E$, which can be either activities or artifacts. In both cases, an element $e \in E$ features two types of associations, i.e., to its prerequisites $\textit{pre}(e)$ and the elements $\textit{post}(e)$, for which it is a prerequisite. The elements and their associations we propose for our ML development process $\mathfrak{D}$ can be seen in Figure~\ref{fig:ml_pipeline}.

When $\mathfrak{D}$ is executed, each of its elements can assume a single state $s^{(t)} \in \{\textit{inactive}, \textit{active}, \textit{done}\} = S$ for each given point in time $t$. We thus define an instance $\mathcal{D}$ of the development process $\mathfrak{D}$ as a set of states $S$ alongside a time line, i.e., a sequence of time points $t \in [t_{\textit{start}},t_{\textit{end}}] \subseteq \mathbb{N}$: At each point in time $t$ the process in its current state $\mathcal{D}^{(t)} : E \to S$ maps each element to one of the states mentioned before. At time point $t_{\textit{start}}$ that mapping is initialized to $D^{t_{\textit{start}}}(e) = \textit{inactive}$ for all $e$. Then, for every element $e$ so that $\forall e' \in \textit{pre}(e): \mathcal{D}^{(t)} = \textit{active}$ we can assign $\mathcal{D}^{(t+1)}(e) = \textit{active}$. After an element has been $\textit{active}$ for at least one time step, it might switch to $\textit{done}$. Note that when activities are \textit{active}, we imagine them being executed by developers or automated procedures, and when artifacts are active, we imagine them being used in some sort of activity. Further note that we only need prerequisites  to be $\textit{active}$ and not $\textit{done}$ as we assume required artifacts to further change alongside the procedures that use them, which is a deviation from what classic waterfall processes might define. Still, an artifact might be $\textit{done}$ immediately after starting all succeeding elements.

One of the main advantages of our formal process model is that it allows arbitrary feedback loops, which means that at any given point in time me may decide to redo any amount of elements in the process as long as we do that in sync. Formally, an element $e$ with $\mathcal{D}^{(t)}(e) \in \{\textit{active},\textit{done}\}$ may switch to $\textit{active}$ or $\textit{inactive}$ as long as all elements $e' \in \textit{post}(e)$ switch to $\textit{active}$ or $\textit{inactive}$. Note that at no point in time may an element $e$ be $\textit{active}$ without all its prerequisites $\textit{pre}(e)$ being $\textit{active}$ or $\textit{done}$. Feedback loops allow us to capture a more dynamic development process while still maintaining a hierarchy of dependencies.

For any instance $\mathcal{D}$ given as a sequence of per-element states, we can now check if $\mathcal{D}$ adheres to the process $\mathfrak{D}$, i.e., $\mathfrak{D} \models \mathcal{D}$. Furthermore, we can reason about the progress of the development using common logics such as LTL~\cite{km08}. We can also use LTL to postulate further constraints on development instances, e.g. the property that the process will eventually produce the desired ML model: $\lozenge \mathcal{D}(\text{``Adapted ML Model''}) = \textit{done}$. Yet, we want to emphasize that other elements $e$ with $\textit{post}(e) = \emptyset$ like the quality seals should be considered equally important results of a development instance.

The immediate expected benefit of reasoning about instances of development processes might be the verification of tools and workflows (``Team A always adheres to the defined process.''), the formulation of clearer process goals (``An instance of a $\textit{done}$ Factory Quality Seal shall occur before $t=100$.''), or the extraction of a better understanding of dependencies and correlations (``Any time the Target Definition started after Data Selection, development was successful.''). But we also see further opportunities in the connection of reasoning about the process and reasoning about the product. To this end, we pay special attention to artifacts regarded as Logical Statements. Ideally, we could define requirements on the development process that reflect in certain guarantees on the software product. Using a more potent logic, e.g., we might be able to formulate a Factory Quality Seal that reads ``The Basic ML Model has been trained successfully at least 3 times in a row with unchanged parameters.''. If we incorporate roles into the model, which is left for future work at the moment, we might even be able to state that ``Data Selection was performed by at least 2 developers with at least 10 years experience between them.'', which naturally might also be part of the specification. Such quality assurance is common in engineering disciplines that are used to never being able to rely on formal proofs or extensive testing. Developing ML software is often more akin to these disciplines, which is why we regard reasoning about the process in addition to the product as so important.

\section{Summary and Outlook}\label{sec:discussion}
So far, we provided an overview of related work and challenges regarding SE for (self-) adaptive and ML systems.
To tackle the challenges, we defined a process model, provided a proof of concept and a formalization.
Now, we conclude by summarizing its strengths and limitations and point to future work.

Through our process model, we hope to close the gaps between existing SD process models and specialized ML training procedures.
It is not restricted to certain ML techniques, i.e. it can be used for Supervised (SL), Unsupervised (UL) and Reinforcement Learning (RL).
Simulated domains or environments for RL can be used analogously to data-sets for SL and UL. 
However, we focus on the life-cycle of the ML model, thus detailed data management activities upfront are omitted.
How to construct data sets such that they correspond to certain requirements is a challenging problem on its own~\cite{pie+21} and not (yet) covered by our approach.
Yet, we believe that data management activities can be integrated straight forward. 
Feedback loops and V\&V activities ensuring data quality can be added similarly to how we used them to ensure quality of the ML model.

Practically, our process model allows to formulate ML SD as a (meta-) optimization problem.
Having human experts tailor the SD processes to the problem domain and algorithm at hand neither scales indefinitely, nor may be optimal for less well understood problem scenarios.
There is a clear trend towards automation in SD with a parallel emergence of powerful optimization techniques in form of ML.
Applying these methods not only on the problem level, but also on SD processes level through sequential decision making algorithms, e.g. RL or Genetic Algorithms, could enable significant progress.
Further cosidering the conceptual overlap of \textit{Ensembles}~\cite{hkp+15} and \textit{Cooperative Multi-Agent RL}~\cite{pga+20,rpm+22}, automated ML seems suitable to realize self-adaptive, autonomous systems.
Vice versa, ML should consider best practices from autonomous computing~\cite{whk+15} on how to handle existing knowledge and how to control automated feedback loops.

Our integration of V\&V acts as a quality gate when transitioning to deployment and to operation.
Depending on the situation, we suggest to use evolutionary or learning methods~\cite{gsp+20,pga+20} or Monte Carlo Based Statistical Model Checking~\cite{pmt20} for testing.
Numerical valuations~\cite{fgs19} can distinguish systems that “barely” satisfy a specification from those that satisfy it in a robust manner.
Still, we rely on a top-level specification, provided in human language, which we assume to change as the SD process progresses.
The open question here is whether we can systematically define the initial specification in a way that ensures that the on-site contract and the SLA will be met.

Next, it should be possible to formulate a consistent mathematical representation of our process model, e.g. through LTL~\cite{km08}.
We plan to tackle this next since it would allow a validation of ML process instances that were created through meta-optimization.
And finally, a key assumption is that a validation of ML processes leads to better ML software.
As we also could not yet engineer an ML component's full life-cycle with the methodology proposed in this paper, both could be combined in future work.




\bibliographystyle{IEEEtranN} 
\bibliography{paper}

\end{document}